# A New Analytical Solution to the Relativistic Polytropic Fluid Spheres

M. I. Nouh[1,3] and A. S. Saad[2,3]

[1] *Physics Dep., Faculty of Science, Northern Border University, Arar, Saudi Arabia*

[2] *Mathematics Dep., Preparatory Year, Qassim University, Qassim, Saudi Arabia*

[3] *National Research Institute of Astronomy and Geophysics, 11421, Helwan, Cairo, Egypt*

**Abstract:**

This paper introduces an accelerated power series solution for Tolman-Oppenheimer-Volkoff (TOV) equation, which represents the relativistic polytropic fluid spheres. We constructed a recurrence relation for the coefficients $a_k$ in the power series expansion $\theta(\xi) = \sum a_k \xi^k$ of the solution of TOV equation. For the range of the polytropic index $0 \leq n \leq 0.5$, the series converges for all values of the relativistic parameters $\sigma$, but it diverges for larger polytropic index. To accelerate the convergence radii of the series, we first used Padé approximation. It is found that the series is converged for the range $0 \leq n \leq 1.5$ for all values of $\sigma$. For $n > 1.5$, the series diverges except for some values of $\sigma$. To improve the convergence radii of the series, we used a combination of two techniques Euler-Abel transformation and Padé approximation. The new transformed series converges everywhere for the range of the polytropic index $0 \leq n \leq 3$. Comparison between the results obtained by the proposed accelerating scheme presented here and the numerical one, revealed good agreement with maximum relative error is of order $10^{-3}$.

**Kew words:** Relativistic fluid sphere, Lane-Emden function, analytical solution, accelerating scheme.

# 1. Introduction

Relativistic effects play an important role in many stellar configurations such as, white dwarfs, neutron stars, black holes, supermassive stars, and star clusters, Synge (1934), Walker (1936), Einstein (1939), Sen and Roy (1954) and Sharma (1988).

Relativistic study of the polytropic equation of state has been done by Tooper (1964). Tooper derived two non linear differential equations analogue to the non-relativistic Lane Emden equation (Tolman-Oppenheimer-Volkoff [TOV] equation ).

The highly non-linear differential equations like TOV equation are usually solved by numerical integration techniques. Ferrari et al. (2007) and Linares et al. (2004), solve numerically the two first order differential equations obtained by Tooper and investigated the effect of increasing a specific relativistic parameter on the polytropes with indices n=1.5 and n=3 respectively.

Undoubtedly true that, the numerical integration techniques, can provide very accurate models. But certainly, if full analytical formulae are established, they definitely become valuable for obtaining models with desired accuracy, Sharma (1981). Moreover, these analytical formulae usually offer much deeper insight into the nature of a model as compared to numerical integration. On the other hand, in the absence of a closed analytical solution of a given differential equation, the power series solution can serve as the analytical representation of its solution, Nouh et al. (2003).

Analytical solutions of TOV equation has been done by Topper (1964) for the polytropic index $n = 0$ only. Sharma (1981) presented an approximate analytical solution for the polytropic indices $n = 0, 1.0\ (0.5)\ 3.0$ and illustrates his results for the relativistic parameters $\sigma = 0\ (0.002)\ 0.04$.

In the present paper, we introduce a new analytic solution to the relativistic fluid sphere. The proposed scheme depends on accelerating the power series solution of the relativistic Emden's function using a combination of two different transformations. Our calculations cover the all possible values of the relativistic parameter ($\sigma$).

The structure of the paper is as follows. In section 2, TOV equation is discussed, section 3 is devoted to the power series solution of TOV equation. The accelerating technique is presented in section 4. The results are summarized in section 5.

## 2. The Relativistic Polytropic Fluid Sphere

For hydrostatic equilibrium stars, Tolman-Oppenheimer-Volkoff (TOV) equation has the form

$$\frac{dP}{dr} = -\frac{G\,\varepsilon(r)\,m(r)}{c^2\,r^2}\left[1+\frac{P(r)}{\varepsilon(r)}\right]\left[1+\frac{4\,\pi^3\,P(r)}{m(r)\,c^2}\right]\left[1-\frac{2G\,m(r)}{c^2\,r}\right]^{-1} \quad (1)$$

The above equation is an extension of the Newtonian formalism with relativistic correction. The equation of state for a polytropic star is $P = k\rho^{1+\frac{1}{n}}$, where $n$ is the polytropic index. Tooper (1964) has shown that the TOV equation together with the mass conservation equation have the form

$$\xi^2 \frac{d\theta}{d\xi}\frac{1-2\sigma\,(n+1)\,\upsilon/\xi}{1+\sigma\,\theta} + \upsilon + \sigma\,\xi\,\theta\frac{d\upsilon}{d\xi} = 0, \quad (2)$$

and $\quad \dfrac{d\upsilon}{d\xi} = \xi^2 \theta^n,$ \quad (3)

with the initial conditions

$\theta(0) = 1, \quad \upsilon(0)=0,$ \quad (4)

where

$$\theta = \rho/\rho_c, \quad \xi = rA, \quad \upsilon = \frac{A^3 m(r)}{4\pi\rho_c}, \quad A = \left(\frac{4\pi G\rho_c}{\sigma(n+1)c^2}\right)^{1/2}, \quad \sigma = \frac{P_c}{\rho_c\,c^2}. \quad (5)$$

$\sigma$ is the relativistic parameter can be related to the sound velocity in the fluid, that is because the sound velocity is given by $v_s^2 = \dfrac{dP}{d\rho}$ in an adiabatic expression. In Equations (5) $\theta$, $\xi$ and $\upsilon$ are dimensionless parameter, while $A$ is a constant.

If the pressure is much smaller than the energy density at the center of a star (i.e. $\sigma$ tends to zero), then we turn to the well known Lane-Emden equation for Newtonian polytropic stars,

$$\frac{1}{\xi^2}\frac{d}{d\xi}\left(\xi^2 \frac{d\theta}{d\xi}\right) + \theta^n = 0 . \tag{6}$$

## 3. Solution of TOV Equation

### 3.1. Operations on Formal Power Series

Algebraic operations on power series can be applied to generate new power series.

- **Multiplication of Two Power Series**

The product of two series is given by

$$\left(\sum_{n=0}^{\infty} a_n x^n\right)\left(\sum_{n=0}^{\infty} b_n x^n\right) = \sum_{n=0}^{\infty} c_n x^n ,$$

where,

$$c_n = \sum_{k=0}^{\infty} a_k b_{n-k} .$$

- **Power Series raised to Powers**

If $n$ is a natural number, then

$$\left(\sum_{k=0}^{\infty} a_k x^k\right)^n = \sum_{k=0}^{\infty} c_k x^k ,$$

where,

$$c_0 = a_0^n ,$$

$$c_m = \frac{1}{m a_0} \sum_{k=1}^{m} (kn - m + k) a_k c_{m-k} .$$

## 3.2. Power Series Solution

Equation (2) can be written in the form

$$\xi^2 \frac{d\theta}{d\xi} - 2\sigma (n+1) \xi \upsilon \frac{d\theta}{d\xi} + \upsilon + \sigma \upsilon \theta + \sigma \xi \theta \frac{d\upsilon}{d\xi} + \sigma^2 \xi \theta^2 \frac{d\upsilon}{d\xi} = 0. \tag{7}$$

Consider a series expansion, near the origin, of the form

$$\theta(\xi) = 1 + \sum_{k=1}^{\infty} a_k \xi^{2k}, \tag{8}$$

that satisfies the initial conditions (4). Then

$$\xi^2 \frac{d\theta}{d\xi} = \sum_{k=1}^{\infty} 2(k+1) \, a_{k+1} \, \xi^{2k+3}. \tag{9}$$

With the help of the algebraic operations on series mentioned in Sec. 3.1, we get

$$\theta^n = \sum_{k=0}^{\infty} \alpha_k \xi^{2k}, \tag{10}$$

where

$$\alpha_0 = a_0^n,$$

$$\alpha_k = \frac{1}{ka_0} \sum_{i=1}^{k} (n \, i - k + i) \, a_i \, \alpha_{k-i}, \quad k \geq 1.$$

Equation (3) can be written in the form

$$\upsilon = \sum_{k=0}^{\infty} \frac{\alpha_k}{(2k+3)} \xi^{2k+3}. \tag{11}$$

Using Equations (9) and (11), we have

$$\xi \frac{d\theta}{d\xi} \upsilon = \sum_{k=0}^{\infty} f_k \, \xi^{2k+2} \sum_{k=0}^{\infty} g_k \xi^{2k+3},$$

where

$$f_k = 2(k+1) \, a_{k+1}, \quad g_k = \frac{\alpha_k}{(2k+3)}.$$

Using the formula of multiplication of two series, yields

$$\xi \frac{d\theta}{d\xi} \upsilon = \sum_{k=0}^{\infty} \gamma_k \, \xi^{2k+5}, \tag{12}$$

where

$$\gamma_k = \sum_{i=0}^{k} f_i \, g_{k-i},$$

$$f_i = 2(i+1)\, a_{i+1},\ g_i = \frac{\alpha_i}{(2i+3)}.$$

From Equations (8) and (11), we obtain

$$\upsilon\, \theta = \sum_{k=0}^{\infty} \eta_k \, \xi^{2k+3}, \qquad \eta_k = \sum_{i=0}^{k} a_i \, g_{k-i}. \tag{13}$$

The usage of Equations (3), (8) and (10) generates

$$\xi\, \theta \frac{d\upsilon}{d\xi} = \sum_{k=0}^{\infty} \beta_k \, \xi^{2k+3}, \qquad \beta_k = \sum_{i=0}^{k} a_i \, \alpha_{k-i}. \tag{14}$$

Equation (14) gives us

$$\xi\, \theta^2 \frac{d\upsilon}{d\xi} = \sum_{k=0}^{\infty} \zeta_k \, \xi^{2k+3}, \qquad \zeta_k = \sum_{i=0}^{k} a_i \, \beta_{k-i}. \tag{15}$$

Substitution of Equations (9) to (15) in Equation (7), evaluates

$$\sum_{k=0}^{\infty} 2(k+1)\, a_{k+1}\, \xi^{2k+3} - 2\sigma(n+1)\, \gamma_k\, \xi^{2k+5} +$$

$$+ \frac{\alpha_k}{(2k+3)}\, \xi^{2k+3} + \sigma\, \eta_k\, \xi^{2k+3} + \sigma\, \beta_k\, \xi^{2k+3} + \tag{16}$$

$$+ \sigma^2\, \zeta_k\, \xi^{2k+3} = 0$$

Equating the like powers of $\xi$ in both sides would determine the coefficients $a_k$ as follows

$$a_{k+1} = \frac{\sigma}{2(k+1)}(2(n+1)\,\gamma_{k-1} - \eta_k - \beta_k + \sigma\, \zeta_k) - \frac{\alpha_k}{2(k+1)(2k+3)},\ k \geq 1 \tag{17}$$

where

$$\gamma_{k-1} = \sum_{i=0}^{k-1} f_i\, g_{k-i-1}, \qquad \eta_k = \sum_{i=0}^{k} a_i\, g_{k-i},$$

$$\beta_k = \sum_{i=0}^{k} a_i\, \alpha_{k-i}, \qquad \zeta_k = \sum_{i=0}^{k} a_i\, \beta_{k-i},$$

$$\alpha_k = \frac{1}{k\, a_0} \sum_{i=1}^{k} (n\, i - k + i)\, a_i\, \alpha_{k-i}, \quad k \geq 1$$

$$\alpha_0 = a_0^n, \quad a_0 = 1.$$

If $\sigma = 0$, we get the coefficients of Emden solution. To get $a_1$, put $k = 0$ in Equation (16), then equate the coefficients of the same power in both sides yields

$$a_1 = -\frac{1}{6} - \frac{\sigma}{2}\left(\sigma + \frac{4}{3}\right).$$

When $\sigma = 0$, then $a_1 = -\frac{1}{6}$ is the first series coefficient in case of Lane-Emden solution.

Tables (1), (2) and (3) show the radius of convergence, $\xi_1(A)$, for **n=0.5, n=1.5** and **n=3** compared with the numerical value, $\xi_1(N)$. Here, **m** is the number of the terms in the series, Equation (17). To obtain reliable comparison between the analytical solution and the numerical one, we integrated Equation (2) and (3) using Runge-Kutta method. A Mathematica routine is elaborated to determine the zeros of TOV equation at different polytropic index **n** and relativistic parameter $\sigma$ (the maximum value of $\sigma$ is given by $\sigma_{max} = n/(n+1)$, where the sound velocity must be smaller than the speed of light). The integrations were started at initial values $\xi = 0$, $\theta = 1$, and $\upsilon = 0$ and proceeded forward using step size $\Delta \xi$. The zero of the function $\theta$, $\xi_1$, is determined by integrating until a negative value of $\theta$ is obtained. Then, small step size $\Delta \xi$ is used to give more accurate results.

**Table (1): Radii of the convergence of $\theta(\xi)$ for n=0.5.**

| $\sigma$ | M | $\xi_1(N)$ | $\xi_1(A)$ |
|---|---|---|---|
| 0.1 | 80 | 2.2899 | 2.2899 |
| 0.2 | 80 | 2.0008 | 2.0008 |
| 0.3 | 100 | 1.8013 | 1.8013 |

**Table (2): Radii of the convergence of $\theta(\xi)$ (Equation (10)) for n=1.5.**

| σ | M | $\xi_1(N)$ | $\xi_1(A)$ |
|---|---|---|---|
| 0.1 | 100 | 3.0384 | 2.5440 |
| 0.2 | 100 | 2.6993 | 2.0824 |
| 0.3 | 200 | 2.4930 | 1.7954 |
| 0.4 | 300 | 2.3610 | 1.5976 |
| 0.5 | 300 | 2.2749 | 1.4965 |
| 0.6 | 400 | 2.2192 | 1.5934 |

**Table (3): Radii of the convergence of $\theta(\xi)$ (Equation (10)) for n=3.**

| σ | M | $\xi_1(N)$ | $\xi_1(A)$ |
|---|---|---|---|
| 0.1 | 100 | 6.8258 | 1.90 |
| 0.2 | 100 | 7.9508 | 1.67 |
| 0.3 | 100 | 10.8337 | 1.49 |
| 0.4 | 200 | 17.8197 | 1.31 |
| 0.5 | 200 | 37.2058 | 1.36 |
| 0.6 | 300 | 91.0723 | 1.17 |
| 0.7 | 400 | 162.5832 | 1.80 |
| 0.75 | 400 | 180.4300 | 1.0 |

As it is clear from these tables, the series diverge for all values of σ even when increasing the series terms to 400 terms, the series diverges at small radius, $\xi_1$.

Divergent or slowly convergent series occur abundantly in the mathematical and physical sciences. Accelerating the convergence of the series by transformation has been done by many authors. Examples are, Euler transformation (Euler, 1755) which is specially designed for the alternating series, $\Delta^2$ process which attributed to Aitken (Aitken, 1926) and Wynn's epsilon algorithm (Wynn, 1956). Nouh (2004) showed that, how one can overcome the slowly convergent series not by straightforward application of a single sequence transformation but by a combination of two different transformations.

## 4. The Acceleration Technique

To accelerate the convergence of the series, we followed the scheme developed by Nouh (2004). In the first step, the alternating series is accelerated by Euler–Abel transformation (Demodovich and Maron, 1973).

Let us write

$$\theta(\xi) = a_0 + \xi\ \phi(\xi),$$

Where

$$\phi(\xi) = \sum_{k=0}^{\infty} a_k\ \xi^{k-1} = \sum_{k=1}^{\infty} a_{k+1}\ \xi^k,$$

then

$$(1-\xi)\ \phi(\xi) = \sum_{k=0}^{\infty} a_{k+1}\ \xi^k - \sum_{k=1}^{\infty} a_k\ \xi^k,$$

$$= a_0 + \sum_{k=0}^{\infty} \Delta a_k\ \xi^k,$$

where

$$\Delta a_k = a_{k+1} - a_k,\quad k = 0,1,2,.......$$

are finite differences of the first order of the coefficients $a_k$. Applying the Euler-Abel transformation to the power series $\sum_{k=0}^{\infty}\Delta a_k\ \xi^k$, $p$ times, and after some manipulations we obtain

$$\sum_{k=0}^{\infty} a_k \xi^k = \sum_{i=0}^{\infty} \Delta^i a_0 \frac{\xi^i}{(1-\xi)^{i+1}} + \left(\frac{\xi}{1-\xi}\right)^p \sum_{k=0}^{\infty}\Delta^p a_k \xi^k, \tag{19}$$

where $\Delta^0 a_0 = a_0$. Equation (19) becomes meaningless when $\xi = 1$, so, by setting $\xi = -t$, we obtain the Euler-Abel transformed series as

$$\theta E_n(t) = \sum_{k=0}^{\infty} \Delta^i a_0 \frac{t^i}{(1-t)^{i+1}} + \left(\frac{t}{1-t}\right)^p \sum_{k=0}^{\infty} \Delta^p \left[(-1)^k a_k\right] t^k. \tag{20}$$

Returning to the earlier variable, $\xi$, we obtain

$$\theta E_n(\xi) = \sum_{i=0}^{p-1}(-1)^i \Delta^i a_0 \frac{\xi_i}{(1+\xi)^i} + \left(\frac{\xi}{1+\xi}\right)^p \sum_{k=0}^{\infty}(-1)^{k+p}\left[\Delta^p a_k\right]\xi_k, \qquad (21)$$

where

$$\Delta^p a_k = \Delta^{p-1} a_{k+1} - \Delta^{p-1} a_k$$

Any order difference can be $\Delta^p a_k$ written as linear combination as

$$\Delta^p a_k = \sum_{i=0}^{p}(-1)^{p-i}\binom{p}{i}a_{k+1},$$

where

$$\binom{p}{i} = \frac{p!}{i!(p-i)!}.$$

The second step is to apply Pade approximation to the Euler-Abel transformed series, Equation (21).

Following the above scheme of acceleration, we obtained the results of Table (4) and (5). The designation in the tables are as follows: $\sigma$ is the relativistic parameter, **m** is the number of terms in the original series, **m₁** is the number of terms in the transformed series, **p** is the times number of applying Euler-Abel transformation, $k \times l$ is the order of Pade approximation, $\xi_1(A)$ is the first zero of $\theta E_n(\xi)$ (Equation (21)) derived analytically, $\xi_1(N)$ is the first zero of $\theta E_n(\xi)$ derived numerically and $\varepsilon$ is the relative error such that

$$\varepsilon = |\xi_1(A) - \xi_1(N)|/\xi_1(N).$$

A trial and error procedure is applied to derive the parameters (**m, m₁, p, $k \times l$**) which controlled by the best relative error. We illustrate the results for n=1.5 and n=3. For n=1.5, p=3 for all values of $\sigma$ except at $\sigma = 0.4$, where p=5. The number of series terms (**m** and **m₁**) required to reach the surface of the polytrope are almost increase as $\sigma$ increase. The maximum number of series terms are 40 with maximum relative error 0.0077.

For n=3, the effect of increasing $\sigma$ on **m** and **m₁** is clear, as $\sigma$ goes high **m** and **m₁** also go high. The maximum number of series terms is 100 with maximum relative error 0.004.

**Table (4). Radii of the convergence of $\theta E_n(\xi)$ (Equation (21)) for n=1.5.**

| $\sigma$ | m | m₁ | P | $k \times l$ | $\xi_1$(N) | $\xi_1$(A) | $\varepsilon$ |
|---|---|---|---|---|---|---|---|
| 0.1 | 20 | 20 | 3 | 6,6 | 3.0384 | 3.0429 | 0.0014 |
| 0.2 | 20 | 20 | 3 | 8,8 | 2.6993 | 2.6966 | 0.0010 |
| 0.3 | 12 | 8 | 3 | 6,6 | 2.4930 | 2.4964 | 0.0013 |
| 0.4 | 32 | 26 | 5 | 20,20 | 2.3610 | 2.3664 | 0.0023 |
| 0.5 | 30 | 30 | 3 | 12,12 | 2.2749 | 2.2806 | 0.0025 |
| 0.6 | 40 | 40 | 3 | 12,12 | 2.2192 | 2.2020 | 0.0077 |

**Table (5). Radii of the convergence of $\theta E_n(\xi)$ (Equation (21)) for n=3.**

| $\sigma$ | M | m₁ | P | $k \times l$ | $\xi_1$(N) | $\xi_1$(A) | $\varepsilon$ |
|---|---|---|---|---|---|---|---|
| 0.1 | 20 | 14 | 5 | 9,9 | 6.8258 | 6.8420 | 0.0024 |
| 0.2 | 72 | 70 | 0 | 22,22 | 7.9508 | 7.9831 | 0.0040 |
| 0.3 | 72 | 70 | 2 | 22,22 | 10.8337 | 10.8414 | 0.0008 |
| 0.4 | 70 | 70 | 1 | 22,22 | 17.8197 | 17.800 | 0.0011 |
| 0.5 | 100 | 100 | 5 | 22,22 | 37.2058 | 37.200 | 0.00015 |
| 0.6 | 100 | 100 | 5 | 24,24 | 91.0723 | 91.080 | 0.000084 |
| 0.7 | 100 | 100 | 0 | 6,6 | 162.5832 | 162.200 | 0.0023 |
| 0.75 | 100 | 100 | 1 | 6,6 | 180.4300 | 180.5000 | 0.00038 |

## 5. Discussion and Conclusion

We presented an accelerated power series solution of relativistic fluid sphere. We used two different accelerating methods to improve the radius of convergence of the series, Euler-Abel transformation and Pade approximation. Applying the proposed accelerating scheme, the transformed series is found to converge to the surface of the polytrope. Comparison with the numerical solution revealed good agreement i.e. the maximum relative error is 0.0077.

Some physical parameters could be established using the series solution of $\theta E_n(\xi)$. In Figures (1) and (2), we plot the density profiles of the stellar matter for n=1.5 and n=3 and for different values of $\sigma$ as a function of the radius $R/R$. The lower panel of each figure illustrates the variation of the relative error with $R/R$.

For n=1.5 we see that, when $\sigma$ increase the stellar, the stellar matter density is more concentrated in the center of the star. For $n=3$, the ultra-relativistic case, the effect is much stronger.

Table (6) gives the estimated maximum relative error for the density and mass profiles. The result reflect good agreement between the analytical solution presented here and the numerical one.

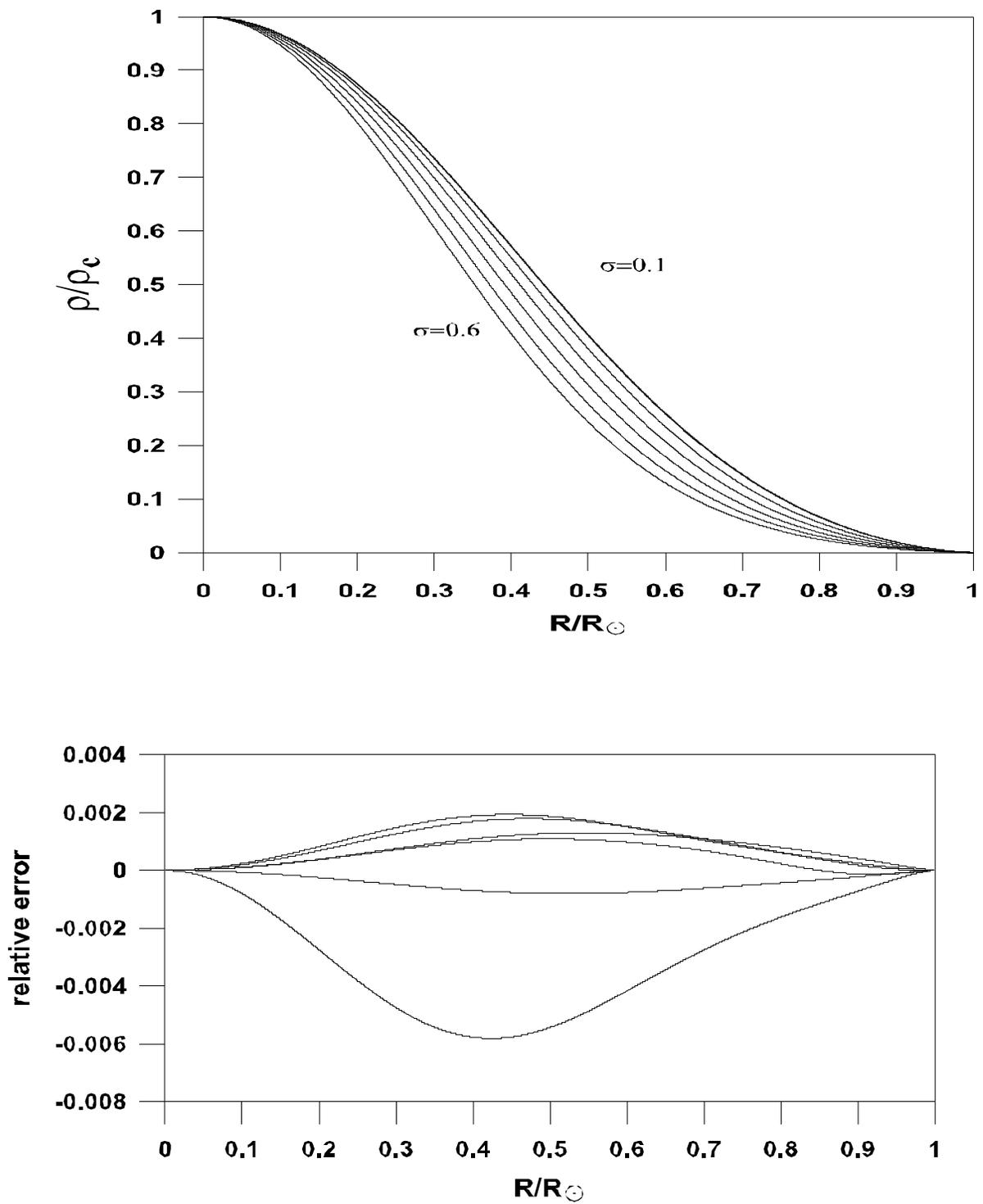

Figure (1). Upper panel: the density profile for n=1.5. Lower panel: the variation of the relative error with radius .

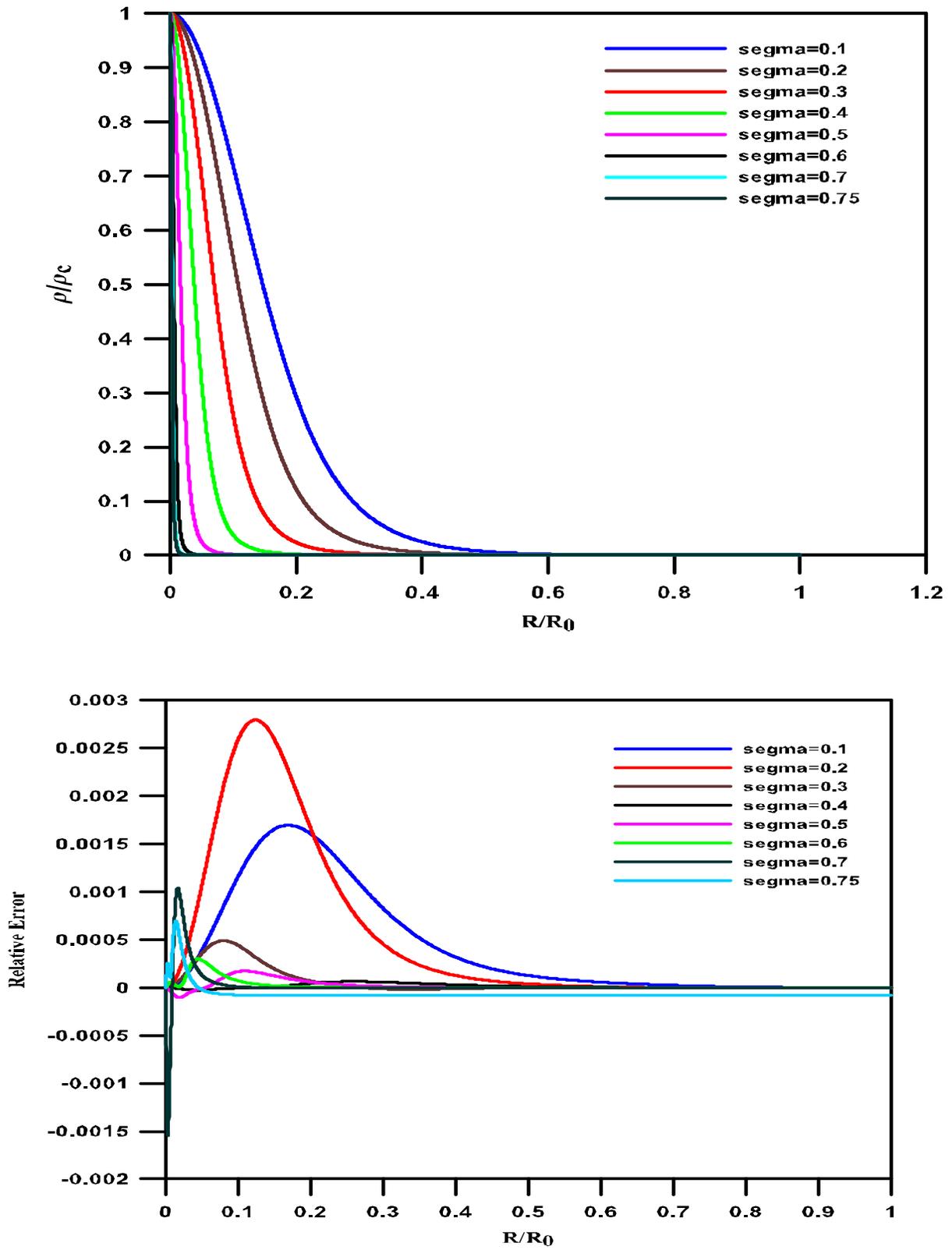

Figure (2). Upper panel: the density profile for n=3. Lower panel: the variation of the relative error with radius .

Table (6). The maximum relative error in the physical parameters at n=3.

| $\sigma$ | Relative error in $\rho/\rho_c$ |
|---|---|
| 0.1 | 0.0012993 |
| 0.2 | 0.0008028 |
| 0.3 | 0.0010595 |
| 0.4 | 0.0016884 |
| 0.5 | 0.0017333 |
| 0.6 | 0.0049308 |